\documentclass[longauth]{aa}  
\usepackage{graphicx}
\usepackage{txfonts}
\usepackage{natbib}
\bibpunct{(}{)}{,}{a}{}{,}
\def\d {\phantom{$0$}}

\def\Mspy   {\ifmmode {M_{\odot} {\rm yr}^{-1}} \else $M_{\odot}$~yr$^{-1}$\fi}
\def\Mdot   {\ifmmode {\dot M} \else $\dot M$\fi}
\def\as     {\ifmmode {\rlap.}$\,$''$\,$\! \else ${\rlap.}$\,$''$\,$\!$\fi}
\def\decsec  {\ifmmode {\rlap.}$\,$^{s}$\,$\! \else ${\rlap.}$\,$^{s}$\,$\!$\fi}\def\decs  {\ifmmode {\rlap.}$\,$^{s}$\,$\! \else ${\rlap.}$\,$^{s}$\,$\!$\fi}
\newcommand{\nhhh}{\mbox{NH$_{3}$}}

\newcommand{\kms}{\mbox{km~s$^{-1}$}}

\begin{document}
\title{\textit{Herschel}/HIFI deepens the circumstellar \nhhh\ enigma\thanks{\textit{Herschel} is an ESA space
observatory with science instruments provided by European-led
Principal Investigator consortia and with important participation
from NASA.}}

 \subtitle{}

 \author{K. M. Menten
 \inst{1}
 \and
 F. Wyrowski
 \inst{1}
 \and
J. Alcolea
        \inst{2}
\and
E. De Beck
\inst{3}
        \and
L. Decin
\inst{3,4}
\and
A. P. Marston
\inst{5}
\and          
V. Bujarrabal
        \inst{6}
        \and
J. Cernicharo
\inst{7}
\and
C. Dominik
\inst{5,8}
\and
K. Justtanont
\inst{9}
\and
A. de Koter
\inst{5,10}
\and
G. Melnick
\inst{11}
\and
D. A. Neufeld
\inst{12}
\and
H. Olofsson
\inst{9,13}
\and
P.  Planesas
\inst{6,15}
   \and 
M. Schmidt
\inst{14}
\and
F. L. Sch\"oier
\inst{9}
   \and 
R. Szczerba
\inst{14}
\and
D. Teyssier
\inst{5}
\and
L. B. F. M. Waters
\inst{4,3}
\and
K. Edwards                    
\inst{16,17}
\and
M. Olberg 
\inst{9,17}                 
\and 
T. G. Phillips
\inst{18}         
\and
P. Morris                   
\inst{19}
\and
M. Salez
\inst{20,21}
\and
E. Caux  
\inst{22,23}
        }


 \institute{   
Max-Planck-Institut f{\"u}r Radioastronomie, Auf dem H{\"u}gel 69,
D-53121 Bonn, Germany 
\email{kmenten@mpifr.de}
\and
Observatorio Astron\'omico Nacional (IGN), Alfonso XII N$^{\circ}$3,
            E--28014 Madrid, Spain  
\and 
Instituut voor Sterrenkunde, Katholieke Universiteit Leuven, Celestijnenlaan 200D, 3001
Leuven, Belgium
\and 
Sterrenkundig Instituut Anton Pannekoek, University of Amsterdam,
Science Park 904, NL-1098 Amsterdam, The Netherlands
\and
European Space Astronomy Centre, ESA, P.O. Box 78, E--28691
Villanueva de la Ca\~nada, Madrid, Spain
\and
Observatorio Astron\'omico Nacional (IGN), Ap 112, E--28803  Alcal\'a de Henares, Spain
\and
CAB, INTA-CSIC, Ctra de Torrej\'on a Ajalvir, km 4, E--28850 Torrej\'on de Ardoz, Madrid, Spain
\and 
Department of Astrophysics/IMAPP, Radboud University, Nijmegen, The Netherlands
\and
Onsala Space Observatory, Dept. of Radio and Space Science, Chalmers  
University of Technology, SE--43992 Onsala, Sweden
\and
Astronomical Institute, Utrecht University, Princetonplein 5, 3584 CC Utrecht, The Netherlands 
\and
Harvard-Smithsonian Center for Astrophysics, 60 Garden Street, Cambridge, MA 02138, USA
\and
The Johns Hopkins University, 3400 North Charles St, Baltimore, MD  21218, USA
\and
Department of Astronomy, AlbaNova University Center, Stockholm University, SE--10691 Stockholm, Sweden
\and
N. Copernicus Astronomical Center, Rabia{\'n}ska 8, 87-100 Toru{\'n}, Poland
\and
Joint ALMA Observatory, El Golf 40, Las Condes, Santiago, Chile
\and
Department of Physics and Astronomy, University of Waterloo, Waterloo, ON Canada N2L 3G1
\and
SRON Netherlands Institute for Space Research, Landleven 12, 9747 AD Groningen, The Netherlands
\and
California Institute of Technology, Cahill Center for Astronomy and Astrophysics 301-17, 
Pasadena, CA 91125 USA
\and
Infrared Processing and Analysis Center, California Institute of Technology, MS 100-22, Pasadena, CA 91125, USA
\and
Laboratoire d'Etudes du Rayonnement et de la Mati\`ere en 
Astrophysique, UMR 8112  CNRS/INSU, OP, ENS, UPMC, UCP, Paris, France 
\and
LERMA, Observatoire de Paris, 61 avenue de l'Observatoire, 75014 Paris, France
\and
Institute Centre d'etude Spatiale des Rayonnements, Universite de Toulouse [UPS], 31062 Toulouse Cedex 9, France 
\and
CNRS/INSU, UMR 5187, 9 avenue du Colonel Roche, 31028 Toulouse Cedex 4, 
France
}

 \date{Received ... ; accepted}
\titlerunning{Ammonia in O-rich evolved stars}
\abstract
{Circumstellar envelopes (CSEs) of a  variety of evolved stars have been found 
to contain ammonia (\nhhh) in amounts that exceed predictions from conventional 
chemical models by many orders of magnitude.} 
{The observations reported here were performed in order  to better constrain the \nhhh\ abundance in 
the CSEs of four, quite diverse, oxygen-rich stars using the \nhhh\ ortho $J_K = 1_0 - 0_0$  ground-state line.} 
{We used the Heterodyne Instrument for the Far Infrared aboard \textit{Herschel}
to observe the \nhhh\ $J_K = 1_0 - 0_0$ transition near 572.5\  GHz, simultaneously with the ortho-H$_2$O $J_{K_a,K_c} = 1_{1,0} - 1_{0,1}$ transition, toward VY CMa, OH 26.5+0.6, IRC+10420, and IK Tau. We conducted non-LTE radiative transfer modeling 
with the goal to derive the \nhhh\ abundance in these objects'  CSEs. For the last two stars, Very Large Array imaging of \nhhh\ radio-wavelength inversion lines were used to provide further constraints, particularly on the spatial extent of the \nhhh-emitting regions.}   
{We find remarkably strong \nhhh\  emission in all of our objects with the \nhhh\ line intensities rivaling those of the ground state H$_2$O line. The  \nhhh\ abundances relative to H$_2$ are very high and range from $2\times10^{-7}$ to $3~\times10^{-6}$   for the objects we have studied. }
{Our observations confirm and even deepen the circumstellar \nhhh\ enigma. While our  radiative transfer modeling does not yield satisfactory fits to the observed line profiles, it does lead to abundance estimates that confirm the very high values found in earlier studies. New ways to tackle this mystery  will include further \textit{Herschel} observations of more \nhhh\ lines and imaging with the Expanded Very Large Array. }
%


 \maketitle
%

\section{Introduction}
Ammonia (NH$_3$) was the first  polyatomic molecule detected in an astronomical 
object \citep{Cheung1968}. It is ubiquitous in dark, dense interstellar cloud cores and 
an eminently  useful thermometer of these regions \citep{WalmsleyUngerechts1983, Danby1988}. This, aided by the easy 
observability of its inversion lines -- many of the astronomically most important ones crowd around 1.3 cm wavelength (24 GHz frequency) -- make  \nhhh\ one of the most frequently observed interstellar molecules \citep{HoTownes1983}.

\nhhh\  has also been detected toward a still limited, but diverse number of 
CSEs around evolved stars, first using infrared (IR) heterodyne absorption spectroscopy toward the high mass-loss asymptotic giant branch (AGB), extreme carbon star CW Leo \citep[= IRC+10216; ][]{Betz1979}. In addition, absorption was found toward a number of oxygen-rich objects that included the long-period variable (LPV)  $o$ Ceti \citep{Betz1985} and the super- or even hyper-luminous objects VY CMa and IRC+10420 \citep{Betz1980, Monnier2000}. The last study finds that around VY CMa the \nhhh\  is forming at $\gtrsim 40$ stellar radii away from the star, where dust formation has well started.

Contemporaneously, several radio inversion lines were detected also toward IRC+10216 \citep{Kwok1981, Bell1982, Nguyen1984}. Later on, high-velocity cm-wavelength \nhhh\ emission plus absorption was found toward the bipolar protoplanetary nebulae (PPNe) CRL 2688 and CRL 618 \citep{Truong1988, Martin-Pintado1992,Truong1996}. Toward CRL 618, P Cygni profiles are observed with a full width at zero power (FWZP) of $\approx 100$~\kms.  \citet{MentenAlcolea1995}  detected high-velocity \nhhh\ radio emission toward IRC+10420, the high mass-loss rate LPV IK Tau, and the PPN OH 231.8+4.2. In the last case, they find high-velocity emission over  $\approx 70$~km~s$^{-1}$ FWZP. Recently, \citet{Hasegawa2006} report and discuss observations of the \nhhh\ line central to the present study, the ortho-NH$_3$ $1_0 - 0_0$ transition, toward IRC+10216 made with the Odin satellite.

One common, surprising  result of \textit{all} the above studies is the exceedingly high \nhhh\ abundances they 
report. Most of them cite values of several times $10^{-7}$ or even $10^{-6}$ relative to molecular hydrogen. These 
numbers are in stark contrast to the results of thermodynamical equilibrium calculations for the atmospheres of cool 
stars, which predict the production of only  negligible amounts of \nhhh, of order $10^{-12}$ \citep{Tsuji1964}. The 
pioneering study of Tsuji has been confirmed by more recent work 
\citep[see, e.g., ][both for C-rich CSEs]{Lafont1982, CherchneffBarker1992}. Somewhat ad hoc approaches to bringing observations and theory closer 
together involved injecting a significant amount of \nhhh\ in the inner envelope \citep{NejadMillar1988, 
Nercessian1989}.  \citet{WillacyCherchneff1998} include shock chemistry  in their model of IRC+10216, but still only 
produce an abundance of $4 \times 10^{-11}$, at least three orders of magnitude below the value implied by 
observations. As to bona fide shocked regions like PPN outflows, \citet{Morris1987} suggested that, for OH 
231.8+4.2, N$_2$, which binds most of the nitrogen, might be dissociated in the high-velocity gas and that 
the high \nhhh\ abundance might be the result of a series of hydrogenation reactions. Whether this can be confirmed 
by detailed  chemical models   remains to be explored.

For the present study, as described in Sect. \ref{obs}, we observed the \nhhh\  $J_K =  1_0 - 0_0$ ortho ground-state 
transition in O-rich stars of widely different natures and mass loss rates: the high mass-loss LPV IK Tau, the peculiar red 
supergiant VY CMa, the archetypical OH/IR star OH 26.5+0.6, and the hypergiant IRC+10420. 
We chose a receiver setting that allowed simultaneous observations of the $J_{K_a,K_c} = 1_{1,0} - 1_{0,1}$ 
transition of ortho-H$_2$O.
All of these objects have dense CSEs, and \nhhh\ has been previously detected toward all of them but OH 26.5+0.6. In particular, for IK Tau and IRC+10420, single-dish observations of the $(J,K) = (1,1)$ and 
$(2,2)$ inversion lines have been reported by \citet{MentenAlcolea1995}. Moreover,  the emission in these lines has subsequently been imaged with the NRAO Very Large Array (VLA) with a  resolution of a few arcseconds (Menten et al., in prep.; see 
Sect. \ref{inv}). 

The critical density of the cm-wavelength inversion lines is on the order of $10^{4}$~cm$^{-3}$, while that of the sub-mm $1_{0} - 0_{0}$ transition has 
a value  $\sim4$ orders of magnitude higher. Thus, both types of lines should provide complementary 
information on different regions of the envelope. 
In Sect.  \ref{results}, we give a general description of the  sub-mm spectra we obtained with HIFI aboard \textit{Herschel} \citep[][in this volume]{Pilbratt2010}. Thereafter, in Sect.  \ref{modeling}, we present radiative transfer calculations conducted to model the observed line profiles, taking advantage of the constraints from the VLA 
imaging. These lead to \nhhh\ abundance determinations.

\section{\label{obs}Observations}
\subsection{\textit{Herschel}/HIFI submillimeter observations}
The observations were made with
the two orthogonal HIFI receivers available for each band, which in all
cases work in double side-band (DSB) mode \citep[see][in this volume]{deGraauw2010}.
This effectively doubles the instantaneous intermediate frequency (IF) coverage. 
We observed the four stars described above  with a  tuning that, in the upper  sideband, covers the frequency of the $J_K 
= 1_0 - 0_0$ ground state transition of ortho-NH$_3$ at 572.4981 GHz. The tuning was chosen to also cover the frequency of the $J_{K_a,K_c} = 1_{1,0} - 
1_{0,1}$ ground state line of ortho-H$_2$O  at 556.9360 GHz  in HIFI's lower sideband. 
The observations were obtained using the dual-beam-switching (DBS)
mode. In this mode, the HIFI internal steering mirror chops between the
source position and a position believed to be free of emission, which  was certainly the case for our observations. The
telescope then alternately locates the source in either of the
chopped beams, providing a double-difference calibration scheme, which
allows a more efficient cancellation of the residual standing waves in
the spectra. Additional details on this observing mode can be found in
\citet[][in this volume]{Helmich2010}. The double sideband system temperature was $\approx 100$ K, and the calibration uncertainty is estimated to be 10\%. Spectral baselines were excellent. \textit{Herschel}'s beam had a size of $37''$ FWHM at the observing frequency, which is much larger than the \nhhh-emitting regions of all our sources.

The HIFI data shown here were obtained using the wide-band
spectrometer (WBS), which is an acousto-optical spectrometer,
providing a simultaneous coverage of the full instantaneous IF band  in the two available orthogonal receivers, with a (oversampled) channel spacing  of 0.5 MHz (0.27 \kms), about half the effective resolution. Spectra in the figures have been resampled and
smoothed to a channel spacing of $\approx 1.1$~\kms. 

The data were processed with the standard HIFI pipeline using HIPE, and nonstitched Level-2 data were exported using the HiClass tool available in HIPE. Further processing, i.e. blanking spurious signals, first order polynomial baseline removal, stitching of the spectrometer subbands and averaging, was performed in CLASS. Since the quality of the spectra measured in both horizontal and vertical polarization was good, these were averaged to lower the final noise in the spectrum. This approach is justified since polarization is not a concern for the presented molecular-line analysis. All HIFI data were originally calibrated in units of antenna temperature ($T_{A}^*$) and were converted to the main-beam temperature ($T_{\rm{MB}}$) scale according to $T_{\rm{MB}}=T_{A}^*/\eta_{\rm{MB}}$, with the main-beam efficiency $\eta_{\rm{MB}}=0.68$. In all cases we have assumed a side-band gain ratio of one.

\subsection{\label{nh3spec}Ammonia spectroscopy and astrophysics}
The main focus of this letter is on the 
$J_K = 1_0 - 0_0$ line of ortho-NH$_3$. Ammonia microwave spectroscopy has a 
long history \citep[see, e.g., ][]{TownesSchawlow1955, Kukolich1967}. 
Very briefly, because of the possible orientations of the hydrogen spins, two different species of \nhhh\ exist that do not interconvert, ortho-NH$_3$ and para-NH$_3$. Ortho-\nhhh\ assumes states, $J_K$, with $K = 0$ or $3n$, where $n$ is an integer (all H spins parallel) , whereas $K \ne 3$ for para-\nhhh\  (not all H spins parallel). 
The principal quantum numbers $J$ and $K$ correspond to the total angular momentum and its projection on the symmetry axis of the pyramidal molecule.

The temperature corresponding to the energy of the lowest para level ($J_K = 1_1$) is 22 K above that of the lowest ortho level ($J_K = 0_0$). Therefore, for formation in the interstellar gas phase, which involves reactions with high exothermicities, the ortho- to para-NH$_3$ ratio is expected to attain its equilibrium value of unity \citep{Umemoto1999}. This situation is also expected to hold for CSEs, given that the IR studies cited above place the \nhhh\ they observe in the hot medium close to the star.

A high-resolution study of 
the \nhhh\ $1_0 - 0_0$ transition  has very recently been presented by \citet{Cazzoli2009}. 
(Only) its upper state is split into several hyperfine 
structure (hfs) components with the  $\sim 1$--2 MHz splitting resulting from the 
coupling of the quadrupole moment of the N nucleus with the electric field of the 
electrons. Two of these components are further split by magnetic interactions.  The 
mean frequency is 572498.1 MHz and the centroid frequencies of the three main hfs 
groups are all within 2 MHz, corresponding to $\approx 1$ \kms, much less than 
the line widths observed for the targets of this study (see Fig. \ref{nh3plot}).

\subsection{\label{inv}VLA observations of inversion lines}
In Sect.  \ref{modeling} we use data of the  $(J,K) = (1,1)$ and $(2,2)$ inversion lines to constrain our 
models for IK Tau and IRC+10420. For these stars, single-dish observations of those 
lines made with the Effelsberg 100 meter telescope were reported by \citet{MentenAlcolea1995}. 
In addition, to place constraints on the spatial distribution of the \nhhh\ molecules, we have used data 
obtained with the VLA  that will be published separately (Menten et 
al. in prep.). The hfs splitting in the inversion lines in velocity units is much wider than for 
the rotation line. 
However, because both the (1,1) and (2,2) lines are very optically thin, as indicated by the 
spectra and supported by our modeling (see Sect.  \ref{modeling}), any contribution of the hfs 
components will be factors of several weaker than the main hfs component and 
neglected in the modeling. The intensities  of the spectra produced from the VLA images, which were restored with a circular beam of $3\as7$ FWHM, used in that section, are consistent with the published  100 m telescope values.

\section{Results and analysis}
\subsection{\label{results}NH$_3$ versus H$_2$O emission}
In Figs. \ref{nh3plot}, \ref{vlaplot}  and  Table \ref{lineresults} we present the results of our HIFI observations of both the \nhhh\ and the H$_2$O ortho ground state lines, together with the results of our \nhhh\ modeling.
All our observed positions agree to within $2''$ with the stars' 2MASS positions, which themselves have an absolute accuracy of better than $0\as1$ \citep{Cutri2003}.
The determination of the LSR ranges is somewhat subjective and the upper and lower velocities are uncertain by $\sim$ a few \kms\ for weaker lines.
For all entries, the formal error in $\int T_{\rm MB}d$$\varv$ is smaller than 0.1 K~\kms.
For VY CMa we used the higher of the literature mass-loss rate values scaled to the recently measured trigonometric parallax distance, 1100 pc \citep{Choi2008}. For both IK Tau and IRC+10420 the lower and the higher values of $X_{{\rm NH}_3}$ are implied by the cm lines and the submm line, respectively.

Inspecting Fig. \ref{nh3plot} and the table, it is striking to see that the luminosity (integrated intensity) in the \nhhh\ and H$_2$O ground-state lines is of comparable magnitude for all of our objects. The \nhhh/H$_2$O line ratios are $\approx 0.50,  0.30, 0.28,$ and 0.42 for IK Tau, VY CMa, OH 26.5+0.6, and IRC+10420, respectively. One has to keep in mind that H$_2$O is a major  molecular constituent of our CSEs, while even the presence of observable  \nhhh\ emission is completely unexplained!

Another remarkable result is that the velocity ranges covered by the two lines are almost identical, which suggests that the bulk of the material producing the emission for both is similar.  Moreover, for all our targets, both lines'  FWZP values are lower, but comparable to twice the terminal velocity,  implying that both molecules are present in the outer layers of the envelope, where the material has almost been fully accelerated.  Furthermore, we point out the clear self absorption in the blue wing of the H$_2$O lines toward IK\,Tau and VY\,CMa, which proves that the line emitting region covers the \emph{whole} envelope.  Whether this is also true for the \nhhh\  line is a priori  not clear.

\subsection{\label{modeling}Radiative transfer modeling and constraints on abundances}
\begin{table*}
\caption{\label{lineresults}Results of HIFI NH$_3$ and H$_2$O observations and NH$_3$ modeling}
\setlength{\tabcolsep}{0.06cm} 
\centering
\begin{tabular}{lcclcccccccc}
\hline \hline
Object          & $\alpha_{\rm J2000}$ 
                                              & $\delta_{\rm J2000}$ 
                                                                      &$D$&$\Mdot$$^{\rm Lit}$&$\varv$$_\infty$
                                                                                         & $\varv$$_{\rm LSR}$ range 
                                                                                            &$\int T_{\rm MB}d$$\varv$&$\Mdot$$^{\rm Mod}$&$r_{\rm out}$&$\theta_{\rm out}$&$X_{{\rm NH}_3}$\\

                    &                     &                       &(kpc)
                                                                                  
                                                                                   &(\Mspy)                            
                                                                                          &(km~s$^{-1})$   &(km~s$^{-1})$    &(K km~s$^{-1})$ &(\Mspy)&(cm) &($''$)\\
IK Tau          & $03^{\rm h}53^{\rm m}28\decsec8$ 
                                             &$+11^\circ 24' 23''$ 
                                                                     & 0.25\{1\}&$8\times10^{-6}$\{2\}&18 \{2\} & [11.5, 57.2] &  \d5.1 &
                                                                                          $8\times10^{-6}$&$5.0\times10^{15}$& 1.3&(1--3)~$10^{-6}$  \\
                    &                      &                      &&&&         [10.4, 59.5] & 10.3\\
VY CMa       & 07 22 58.3 &$-25$ 46 03 & 1.1 \{3\} &(0.8--3.4)~$10^{-4}$\{4\} &  35 \{4\}& [$-23.7$, 64.3] &15.5 &
								$1.8\times10^{-4}$&$3.5\times10^{16}$&2.1&$3~10^{-6}$\\ 
                    &                     &                       &&&&       [$-34$, 79]    & 51.2\\
OH26.5+0.6& 18 37 32.5 &$-05$ 23 59 & 1.37 \{5\}  &$2.4\times10^{-4}$ \{5\} & 16 \{6\} &  [14.6, 43.2]  &\d0.96&
								 $2.4\times10^{-4}$       &$2.5\times10^{16}$&1.2&$3~10^{-7}$\\
                     &                    &                       &&&&        [15.0, 43.2]  &\d3.4 \\      
IRC+10420 & 19 26 48.1 &  +11   21 17& 5 \{7\}  &$8\times10^{-4}$ \{7\}      &   	38 \{7\}&[39,112]     & \d3.1&
								$2\times10^{-3}$        &$1.5\times10^{17}$      &2.0& (2--5)~$10^{-7}$\\ 
                     &                    &                       &&&&        [42.8,116.4] &\d7.4\\
\hline
\end{tabular}
\tablefoot{
Columns are (from left to right) the object, J2000 right 
ascension and declination of observed position, distance,  mass-loss rate from the literature, terminal expansion velocity (from CO data), FWZP LSR velocity range 
with observed emission, integrated main-beam brightness temperature, as well as mass-loss, outer radius of the NH$_3$ distribution (in cm and in arcseconds),  and   [\nhhh/H$_2$] abundance ratio derived from our modeling. For the $\varv$$_{\rm LSR}$ range and $\int T_{\rm MB}d$$\varv$  for each source, the first and second rows  give the values for the 
\nhhh\ and the H$_2$O lines, respectively. Numbers in braces relate to the following references for distances and mass loss rates: 
\{1\} \citet{Hale1997},
\{2\} \citet{Decin2010}, 
\{3\} \citet{Choi2008},
\{4\} \citet{Decin2006}, 
\{5\} \citet{Justtanont2006}, 
\{6\} \citet{Kemper2003},
\{7\} \citet{trung2009}.
}
\end{table*}

\begin{figure}
\centerline{\resizebox{0.77\hsize}{!}{\includegraphics[angle=0]{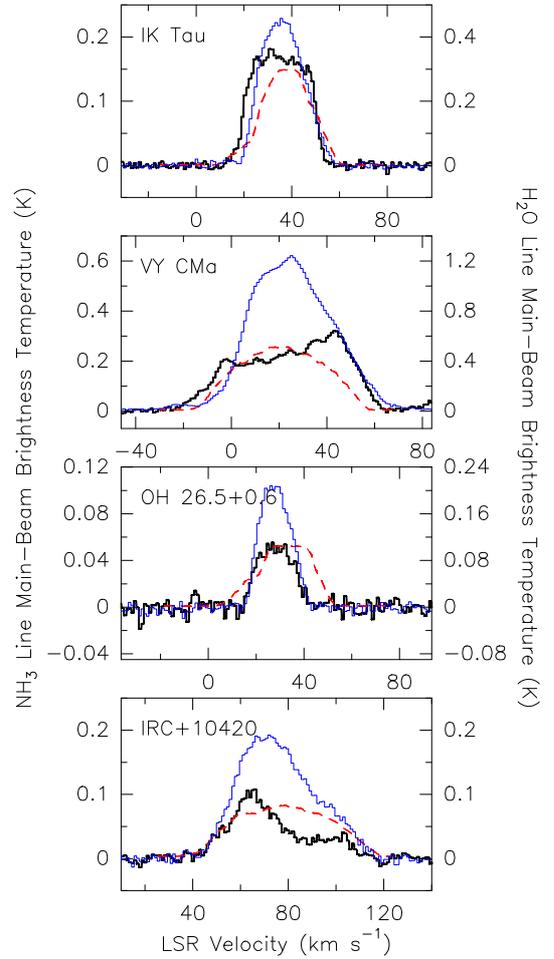}}}
\caption{HIFI spectra compared to radiative transfer model results: The thick black lines show spectra of the  $J_K = 1_0 - 0_0$ transition 
of ortho-NH$_3$ for (top to bottom) IK Tau, VY CMa, OH 26.5+0.6, and IRC+10420. 
The blue lines represent the ortho-H$_2$O   $J_{K_a,K_c} = 1_{1,0} - 1_{0,1}$ 
spectra for the same stars in the same order. The  dashed red lines represent the predictions for the NH$_3$ line resulting from our radiative transfer modeling. The left- and righthand ordinates give 
the main-beam brightness temperature scales for the NH$_3$ and the H$_2$O lines, 
respectively. Except for IRC+10420, they have different ranges.}
\label{nh3plot}
\end{figure}

The \nhhh\ emission of the sources has been modeled with the Monte
Carlo radiative transfer code RATRAN 
developed and described by  \citet{hogerheijde2000}.  For \nhhh, RATRAN uses collision rates calculated by \citet{Danby1988}. Power
laws for the density and temperature were used to describe the
physical structure of the envelope, using as input published values for
the mass loss rate and expansion velocity  (see Table \ref{lineresults}).  
For
IRC+10420, envelope parameters from \citet{trung2009} were used and
for IK Tau, VY CMa, and OH26.5+0.6 we refer to the modeling of \citet{Decin2010},
\citet{Decin2006}, and \citet{Justtanont2006}, respectively.
The VLA data yield an 
extent of $\approx 4''$ for the \nhhh\ emitting region around IRC+10420 and 
$\approx 2\as6$ for that around IK Tau, numbers we use for our modeling.

To fit the submm \nhhh\ lines observed with HIFI, the \nhhh\ abundance
was varied in a first iteration. For the two sources with additional
data from the cm inversion lines, those (para) lines were modeled as
well, using an ortho-to-para ratio of 1, appropriate for formation of
\nhhh\ under high temperatures (see Sect.  \ref{nh3spec}).
Interestingly, this does not lead to a satisfying fit for both the cm
\nhhh\ inversion lines \textit{and} the
submm ground state line. With a fit
adjusted to reproduce the cm lines, the submm line is underestimated by a factor
of 10 for IK Tau. 

Since high densities are needed to excite the submm line, its emission must arise from the inner part of the envelope, further inward than the cm-line emitting region. 
Our modeling suggests densities above a few times $10^6$~ cm$^{-3}$
and temperatures in the 10--100 K range.
This discrepancy between the physical conditions required to produce the observed cm emission, on the one hand, and the submm emission, on the other, is even greater for the very extended but relatively low-density shell
of IRC+10420. To reach densities high enough to excite the \nhhh\
sub-mm line, the mass-loss rate had to be increased to  $2\times10^{-3}$ \Mspy, i.e.,  a factor
3 higher than the value derived by \citet{trung2009} and the inner radius reduced to $5\times 10^{15}$~cm, which is twice the value of the hot inner shell proposed by these authors. Then agreement between the
abundances obtained from the cm and submm lines can be reached within
a factor of 2. Interestingly, De Beck et al. (2010, accepted for publication), derive a value of $3.6\times10^{-3}$ \Mspy\ for IRC+10420's mass loss rate  based on multi-transition modeling of CO.

To reproduce the strong submm \nhhh\ line from VY CMa, the highest mass-loss rate and largest radius from the various shells discussed by
\citet{Decin2006} had to be used, scaled to $D = 1100$~pc (see Sect. \ref{results}).
The outer radius that led to a best fit
for OH26.5+0.6 is $2.5\times 10^{16}$~cm. For this source, the same
temperature profile as for IK Tau was used.

The line profiles produced by our \nhhh\ model calculations are shown in Figs. \ref{nh3plot} and \ref{vlaplot} overlaid on the measured spectra.
The resulting \nhhh\ abundances for the four observed stars 
range from $2\times 10^{-7}$ to  $3\times 10^{-6}$  (see Table \ref{lineresults}). While they are 
in line with circumstellar  \nhhh\ abundances derived from the inversion lines alone and from IR spectroscopy, 
we note, as a caveat, that our model calculations did not consider the possibility of IR pumping of the $1_0 - 0_0$ transition. 
IR pumping of the H$_2$O $1_{1,0}  - 1_{0.1}$ line in IRC+10216's CSE  via various vibrational bands has been investigated by 
\citet{AgundezCernicharo2006}, who found it to be the dominant source of excitation of this high critical density line
over much of the star's outer envelope. Consequently, their modeling suggests an order of magnitude lower H$_2$O abundance 
than invoked earlier from SWAS and Odin observations \citep{Melnick_etal2001, Hasegawa2006}. 

\section{Summary and outlook}
The high critical density of the $J_K = 1_0 - 0_0$ ortho-\nhhh\ line 
allows investigations of \nhhh\ in a new density regime of circumstellar envelopes. However, the remarkably high
abundances we determine for all our objects confirm and significantly strengthen
the finding that this molecule exists in a variety of CSEs at levels not explained by current 
chemical models. Better constraints on the emitting regions will come from HIFI observations 
of a number of \nhhh\ lines, which are actually 
evident in the (heavily spectrally diluted) PACS spectrum of VY CMa \citep{Royer2010}.  
Given their intensities, such observations appear eminently feasible. They would place tight constraints on future  radiative transfer modeling that includes
infrared excitation. Additionally, with its extremely 
wideband new generation digital correlator, the Expanded Very Large Array (EVLA) will allow 
simultaneous imaging of many NH$_3$ inversion lines. Nevertheless, given the weakness of these lines
toward ordinary AGB stars, PPNe being an exception, EVLA observations will be challenging.


\Online
\begin{appendix}
\section{NH$_3$ rotational and inversion line spectra and fits for IK Tau and IRC+10420}
\begin{figure}[h]
\centerline{\resizebox{\hsize}{!}{\includegraphics[angle=0]{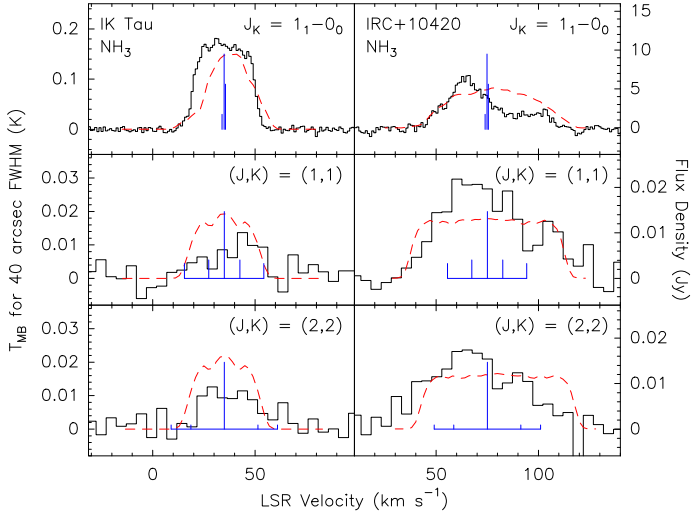}}}
\caption{Spectra of \nhhh\ transitions for IK Tau (left column) and IRC+10420 (right 
column).
Top row: HIFI spectra of the $J_K = 1_0 - 0_0$ ortho-NH$_3$ line. Middle row:  VLA 
spectra of the  $(J,K) = (1,1)$ para-NH$_3$ line.
Bottom row: VLA spectra of the  $(J,K) = (2,2)$ para-NH$_3$ line. The intensity scales 
apply for both sources. The lefthand ordinate gives main-beam brightness temperature  in a $40''$ FWHM 
beam, while the righthand ordinate gives flux density in Jy units. In all  panels the red dashed line represents our best-fit model prediction.
The blue bars give the spacings and theoretical relative intensities of the main groups of hfs components determined by \citet{Cazzoli2009} for the $1_0 - 0_0$ transition and by \citet{Kukolich1967} for the (1,1) and (2,2) inversion lines, respectively.}
\label{vlaplot}
\end{figure}

Fig. \ref{vlaplot} shows, for IK Tau and IRC+10420, the HIFI spectra of the $J_K = 1_0 - 0_0$ ortho-NH$_3$ line and the  spectra of the $(J,K) = (1,1)$ and $(2,2)$ para-NH$_3$ lines produced from our VLA data together with our best fit model predictions. 

\end{appendix}

\begin{acknowledgements}
HIFI has been designed and built by a consortium of institutes and
university departments from across Europe, Canada, and the United States
under the leadership of SRON Netherlands Institute for Space Research,
Groningen, The Netherlands and with major contributions from Germany,
France, and the US.  Consortium members are: Canada: CSA, U.Waterloo;
France: CESR, LAB, LERMA, IRAM; Germany: KOSMA, MPIfR, MPS; Ireland,
NUI Maynooth; Italy: ASI, IFSI-INAF, Osservatorio Astrofisico di
Arcetri- INAF; Netherlands: SRON, TUD; Poland: CAMK, CBK; Spain:
Observatorio Astron\'omico Nacional (IGN), Centro de Astrobiolog\'{\i}a
(CSIC-INTA); Sweden: Chalmers University of Technology - MC2, RSS \&
GARD; Onsala Space Observatory; Swedish National Space Board, Stockholm
University - Stockholm Observatory; Switzerland: ETH Zurich, FHNW; USA:
Caltech, JPL, NHSC. HCSS / HSpot / HIPE is a joint development by the
\textit{Herschel} Science Ground Segment Consortium, consisting of ESA, the NASA
\textit{Herschel} Science Center, and the HIFI, PACS and SPIRE consortia.
This work has been partially supported by the
Spanish MICINN, within the program CONSOLIDER INGENIO 2010, under grant
``Molecular Astrophysics: The \textit{Herschel} and ALMA Era -- ASTROMOL" (ref.:
CSD2009-00038). R. Sz.\ and M. Sch.\ acknowledge support from grant N 203
393334 from the Polish MNiSW. K.J.\ acknowledges the funding from SNSB.
J.C.\ thanks funding from MICINN, grant AYA2009-07304.
\end{acknowledgements}

\end{document}